\documentclass[aoas,MSNbibl,nameyear,rotating,dvips]{arximspdf}
\usepackage{dcolumn}
\usepackage{graphicx}

\doi{10.1214/14-AOAS783} 
\volume{8}
\issue{4}
\pubyear{2014}
\firstpage{2461}
\lastpage{2484}
\docsubty{FLA}
\setattribute{copyright}{owner}{In the Public Domain}

\makeatletter
\newcolumntype{d}[1]{D{.}{.}{#1}}
\renewcommand{\mid}{|}
\newcommand{\hunt}{\mathrm{hunt}}
\makeatother

\begin{document}
\begin{frontmatter}

\title{Probit models for capture--recapture data subject to imperfect
detection, individual heterogeneity and misidentification\thanksref{T1}}
\runtitle{Probit capture--recapture misidentification models}

\begin{aug}
\author[A]{\fnms{Brett T.}~\snm{McClintock}\corref{}\thanksref{m1}\ead[label=e1]{brett.mcclintock@noaa.gov}},
\author[B]{\fnms{Larissa L.}~\snm{Bailey}\thanksref{m2}\ead[label=e2]{larissa.bailey@colostate.edu}},\\
\author[C]{\fnms{Brian P.}~\snm{Dreher}\thanksref{m3}\ead[label=e3]{brian.dreher@state.co.us}}
\and
\author[D]{\fnms{William A.}~\snm{Link}\thanksref{m4}\ead[label=e4]{wlink@usgs.gov}}
\runauthor{McClintock, Bailey, Dreher and Link}
\affiliation{NOAA National Marine Mammal Laboratory\thanksmark{m1},
Colorado State University\thanksmark{m2},
Colorado Parks and Wildlife\thanksmark{m3},\break and
USGS Patuxent Wildlife Research Center\thanksmark{m4}}
\address[A]{B.~T. McClintock\\
National Marine Mammal Laboratory\\
Alaska Fisheries Science Center, NOAA-NMFS\\
7600 Sand Point Way NE\\
Seattle, Washington 98115\\
USA\\
\printead{e1}}
\address[B]{L. L. Bailey\\
Department of Fish, Wildlife,\\
\quad and Conservation Biology\\
Colorado State University\\
Fort Collins, Colorado 80523\\
USA\\
\printead{e2}}
\address[C]{B. P. Dreher\\
Colorado Parks and Wildlife\\
4255 Sinton Road\\
Colorado Springs, Colorado 80907\\
USA\\
\printead{e3}}
\address[D]{W. A. Link\\
USGS Patuxent Wildlife\\
\quad Research Center\\
12100 Beech Forest Road\hspace*{23pt}\\
Laurel, Maryland 20708\\
USA\\
\printead{e4}}
\end{aug}
\thankstext{T1}{Funding for the bear example
was provided in part by Federal Aid in Wildlife Restoration Project
W-147-R-2, Michigan Department of Natural Resources, United States Fish
and Wildlife Service, and Michigan State University.}

\received{\smonth{12} \syear{2013}}
\revised{\smonth{5} \syear{2014}}

%
\begin{abstract}
As noninvasive sampling techniques for animal populations have become
more popular, there has been increasing interest in the development of
capture--recapture models that can accommodate both imperfect detection
and misidentification of individuals (e.g., due to genotyping error).
However, current methods do not allow for individual variation in
parameters, such as detection or survival probability. Here we develop
misidentification models for capture--recapture data that can
simultaneously account for temporal variation, behavioral effects and
individual heterogeneity in parameters. To facilitate Bayesian
inference using our approach, we extend standard probit regression
techniques to latent multinomial models where the dimension and zeros
of the response cannot be observed. We also present a novel
Metropolis--Hastings within Gibbs algorithm for fitting these models
using Markov chain Monte Carlo. Using closed population abundance
models for illustration, we re-visit a DNA capture--recapture population
study of black bears in Michigan, USA and find evidence of
misidentification due to genotyping error, as well as temporal,
behavioral and individual variation in detection probability. We also
estimate a salamander population of known size from laboratory
experiments evaluating the effectiveness of a marking technique
commonly used for amphibians and fish. Our model was able to reliably
estimate the size of this population and provided evidence of
individual heterogeneity in misidentification probability that is
attributable to variable mark quality. Our approach is more
computationally demanding than previously proposed methods, but it
provides the flexibility necessary for a much broader suite of models
to be explored while properly accounting for uncertainty introduced by
misidentification and imperfect detection. In the absence of
misidentification, our probit formulation also provides a convenient
and efficient Gibbs sampler for Bayesian analysis of traditional closed
population capture--recapture data.
\end{abstract}

%
\end{frontmatter}

\section{Introduction}\label{secintro}

Capture--recapture methods are commonly used to estimate demographic
parameters for wildlife [e.g., \citet{WilliamsEtAl2002}] and human
[e.g., \citeauthor{YipEtAl1995a} (\citeyear{YipEtAl1995a,YipEtAl1995b})] populations. Passive (or
``noninvasive'') sampling techniques are becoming more common in
capture--recapture studies, largely because these techniques can be less
expensive and less invasive than the physical capture of animals [e.g.,
\citet{KaranthNichols1998,MackeyEtAl2008,RuellEtAl2009}]. Passive
sampling techniques in capture--recapture studies include the use of
photographs [\citet
{KaranthNichols1998,LangtimmEtAl1998,MackeyEtAl2008}], visual sightings
[e.g., \citet{HallEtAl2001,KauffmanEtAl2003}] or genetic material
[\citet{DreherEtAl2007,RuellEtAl2009}] to individually identify
animals. When individual animals are identifiable by natural or
artificial marks, these techniques can provide information about key
demographic parameters such as abundance, survival and recruitment.
They are therefore very useful for informing management decisions, as
well as for testing ecological or evolutionary hypotheses.

Unfortunately, use of passive sampling techniques in capture--recapture
studies is not entirely without problems. For example, matching
photographs to individuals can be prone to identification error due to
variable image quality [e.g., \citet
{HastingsEtAl2008,LinkEtAl2010,MorrisonEtAl2011,BonnerHolmberg2013,McClintockEtAl2013a}],
and genetic samples (e.g., scat or hair) are susceptible to genotyping
error [e.g., \citet
{LukacsBurnham2005,DreherEtAl2007,WrightEtAl2009}]. Individual
identifications from photographs, visual sightings or genetic samples
are all susceptible to observer recording error. Sampling designs can
also result in differential exposures of individuals to sampling (e.g.,
due to home range behavior or opportunistic sampling). Such individual
heterogeneity in detection probabilities can severely bias estimators
and is a common culprit in the underestimation of abundance in
capture--recapture studies.

\citet{LinkEtAl2010} recently developed a novel approach for the
analysis of capture--recapture data when individual identification
errors occur. This pioneering contribution focused on the closed
population abundance model allowing for temporal variation in
parameters [\citet{Darroch1958,OtisEtAl1978}], and therefore does
not accommodate individual-level variation in parameters, such as
detection [e.g., \citet
{Coull-Agresti1999,FienbergEtAl1999,Pledger2000,Basu-Ebrahimi2001,KingBrooks2008,ManriqueVallier-Fienberg2008}]
or survival [e.g., \citet{Royle2008,GimenezChoqet2010}]
probability. Here, we develop models to simultaneously account for
temporal variation, behavioral response (e.g., trap ``happy'' or
``shy'' effects), individual heterogeneity and misidentification in
capture--recapture analyses. To facilitate Bayesian inference using our
approach, we also extend standard probit regression data augmentation
techniques [e.g., \citet{Albert-Chib1993}] to latent multinomial
models where the dimension and zeros of the response cannot be observed.

\section{Methods}\label{secmethods}
\subsection{Detailed problem description}\label{subsecproblem}
Consider a ``classic'' capture--recapture study, where sampling is
conducted over $T$ sampling occasions and the identity of each animal
is known with certainty when it is observed (i.e., there is no
misidentification). When encounters are simple binary responses and
$T=2$, there are three possible recorded encounter histories for each
individual: ``11'' (encountered on both occasions), ``10'' (encountered
on the first occasion but not the second) and ``01'' (encountered on
the second occasion but not the first). If the encounter history for
animal $i$ is denoted $\mathbf{h}_i$, a classic approach is to assume that
$\mathbf{h}_i$ is a realization from a multinomial process, where the
probability of observing $\mathbf{h}_i$ is a function of unknown
demographic parameters ($\bolds{\theta}$) and (usually nuisance)
parameters related to the observation process ($\bolds{\rho}$). For
example, $\bolds{\theta}$ might consist of survival probabilities and $\bolds
{\rho}$ of detection probabilities. In this case, the number of unique
animals encountered ($n$) is known with certainty, and when
conditioning on first capture, a standard likelihood for
capture--recapture data is proportional to
%
%
\begin{equation}
[ {\mathbf h} \mid{\bolds\theta},{\bolds\rho} ] = \prod_{i=1}^n
\operatorname{Pr}(\mathbf{h}_i|{\bolds\theta},{\bolds\rho}), \label{eqcanonicalL}
\end{equation}
where $ [ {\mathbf h} \mid{\bolds\theta},{\bolds\rho}  ]$ denotes the
conditional distribution for ${\mathbf h}$ given ${\bolds\theta}$ and ${\bolds
\rho}$. We note that ``00'' encounter histories are not observed,
hence, additional modifications to equation (\ref{eqcanonicalL}) are
needed to make inferences about individuals that are never encountered.

In contrast to the preceding scenario, now consider the situation where
individuals may be misidentified. When such errors can occur, three
types of encounters for any of the $T$ sampling occasions are possible.
These include a nonencounter (denoted by ``0''), a correctly identified
encounter (denoted by ``1'') or a misidentified encounter (denoted by
``2''). Misidentified encounters result in ``ghost'' encounter
histories [\citet{Yoshizaki2007,LinkEtAl2010}], and an individual
encountered in $>$1 sampling occasion could therefore yield a number of
possible recorded histories. For example, when presented with the
recorded histories ``10'' and ``01,''\vadjust{\goodbreak} we do not know whether these
observations arose from the same animal seen on both occasions (latent
histories ``12,'' ``21'' or ``22'') or whether it was indeed two
different animals each seen on one occasion (latent histories ``10''
and ``01,'' ``20'' and ``01,'' ``10'' and ``02,'' or ``20'' and
``02''). Under misidentification, encounter histories are not uniquely
associated with animals, so equation (\ref{eqcanonicalL}) is no longer
valid for making inferences about $\bolds{\theta}$ and $\bolds{\rho}$.

Assuming the same misidentification cannot occur more than once (i.e.,
a ghost cannot be detected more than once) and an encounter cannot be
misidentified as a legitimate marked individual, \citet
{LinkEtAl2010} proposed a closed population abundance model allowing
for temporal variation in detection probability under this
misidentification scenario. In the next section, we generalize their
approach to a much broader suite of misidentification models that can
simultaneously accommodate temporal, behavioral and individual effects
on $\bolds{\theta}$ and $\bolds{\rho}$.

\subsection{Accounting for individual heterogeneity and misidentification}\label{subsecaccount}
Consider the marginal likelihood obtained by summing the ``complete
data likelihood'' over all possible values of the latent encounter histories:
%
%
\begin{equation}
[ {\mathbf f} \mid{\bolds\theta},{\bolds\rho} ]=\sum_{{\mathbf h}} [ {
\mathbf h} \mid{\bolds\theta},{\bolds\rho} ] [ {\mathbf f} \mid {\mathbf h},{\bolds\theta},{\bolds\rho} ],
\label{eqLcomplete}
\end{equation}
where ${\mathbf f}$ is a vector of recorded history frequencies indicating
the number of times each of the possible recorded histories was
observed (see Table~\ref{tabdefs} for notation definitions). The
complete data likelihood therefore derives from distributions $  [
{\mathbf h} \mid{\bolds\theta},{\bolds\rho}  ]$ for latent
capture--recapture data and distributions $ [ {\mathbf f} \mid{\mathbf
h},{\bolds\theta},{\bolds\rho}  ]$ describing their conversion to
observed (potentially misidentified) data (see Table~\ref{tabexample}).
We note that this extension is applicable to all sorts of
capture--recapture models [e.g., those reviewed by \citet
{WilliamsEtAl2002}] and could apply to data subject to errors other
than misidentification [e.g., incomplete mark observations sensu
\citet{McClintockEtAl2013b}]. Evaluating equation (\ref{eqLcomplete})
involves a multidimensional summation, thus making maximum likelihood
estimation difficult. \citet{LinkEtAl2010} averted this problem by
adopting a Bayesian perspective and sampling from the posterior
distribution using Markov chain Monte Carlo (MCMC), but their approach
requires the assumption of no individual variation in ${\bolds\theta}$ and~${\bolds\rho}$.
%
%
\begin{table}
\tabcolsep=0pt
\caption{Definitions of parameters, latent variables, data and modeling
constructs used in the latent multinomial model allowing
misidentification with temporal, behavioral and individual-level
variation in parameters. Note that bold symbols represent collections
(vectors) of parameters}\label{tabdefs}
\begin{tabular*}{\tablewidth}{@{\extracolsep{\fill}}@{}lp{270pt}@{}}
\hline
\textbf{Parameters}  & \textbf{Definition}\\
\hline
${\bolds\theta}$ &  Vector of demographic process parameters (e.g., abundance or survival probability). \\[3pt]
$\bolds{\rho}$ &  Vector of observation process parameters (e.g., encounter or misidentification probability). \\[3pt]
$p_{it}$ &  Probability that individual $i$ is encountered at time $t$. \\[3pt]
$\alpha$ &  Probability that an individual, encountered at time $t$, is correctly identified.
\\[6pt]
\textbf{Latent variables} &  \\
\hline
$\mathbf{h}_i$ &  The latent encounter history for individual $i$, $ (h_{i1},h_{i2},\ldots,h_{iT}  )$. \\[3pt]
$h_{it}$ &  Encounter type for the latent encounter history of individual $i$ at time $t$; $h_{it}=0$ represents no encounter, $h_{it}=1$ a correctly identified encounter, and $h_{it}=2$ a misidentified encounter. \\[3pt]
$H_i$ &  Latent encounter history index for individual $i$, such that $H_i=j$ indicates individual $i$ has latent history $j$. For $h_{it} \in \{ 0,1,2  \}$ the $3^T$ possible latent histories are identified by $j=1+\sum_{t=1}^T h_{it} 3^{t-1}$ (see Table~\ref {tabexample}). \\[3pt]
$x_j$ &  Latent frequency of encounter history $j$, where $x_j=\sum_i \mathrm{I} (H_i=j  )$. Note that ${\mathbf x}$ denotes a column vector of such frequencies, for example, ${\mathbf x}=(x_1,x_2,\ldots,x_{3^T})^\prime$ for $h_{it} \in \{ 0,1,2 \}$.
\\[6pt]
\textbf{Data} &  \\
\hline
$T$ &  Number of sampling occasions.\\[3pt]
$f_k$ &  Frequency for recorded (observed) encounter history $k$. Note that ${\mathbf f}$ denotes a column vector of such frequencies, for example, ${\mathbf f}=(f_1,f_2,\ldots,f_{2^T-1})^\prime$ for $\omega_t \in \{0,1  \}$.
\\[6pt]
\textbf{Modeling constructs} &  \\
\hline
$\bolds{\omega}$ &  Recorded encounter history, $ ( \omega_1,\omega_2,\ldots,\omega_T  )$.\\[3pt]
$\omega_{t}$ &  Observation type for a recorded history at time $t$; $\omega_t=0$ represents no detection and $\omega_t=1$ a detection. For $\omega_t \in \{ 0,1  \}$ the
$2^T-1$ possible recorded histories are identified by $k=\sum_{t=1}^T \omega_{t} 2^{t-1}$ (see Table~\ref{tabexample}). \\[3pt]
$C_i$ &  Occasion of first capture for individual $i$. For example, $C_i=3$ if individual $i$ has latent encounter history ${\mathbf h}_i=0021$
$(H_i=46)$. \\
\hline
\end{tabular*}
\end{table}
%
%
\begin{table}
\tabcolsep=0pt
\caption{Latent and recorded histories from marked individual
encounters with $T = 3$ sampling occasions subject to
misidentification. The probability of each latent history for
individual $i$, $\operatorname{Pr}(H_i=j)$, is for a closed population abundance
model, where $p_{it}$ is the probability that individual $i$ is
encountered at time $t$, and $\alpha$ is the probability that an
individual, encountered at time $t$, is correctly identified.
Contributed records column shows the recorded histories $(k)$ arising
from specific latent histories~$(j)$. For example, latent history 25,
``022,'' gives rise to recorded histories ``010'' and ``001'' (for
which $k = 2$ and $4$)}\label{tabexample}
\begin{tabular*}{\tablewidth}{@{\extracolsep{\fill}}@{}lccccc@{}}
\hline
 & \textbf{Latent} &  & \textbf{Contributed} &  & \textbf{Recorded} \\
& \textbf{history} & & \textbf{records} & & \textbf{history}\\
$\bolds{j}$ & $\bolds{({\mathbf{h}}_i)}$ & $\bolds{\operatorname{Pr}(H_i=j)}$& \textbf{($\bolds{k}$ from $\bolds{j}$)} & $\bolds{k}$ & $\bolds{({\omega})}$ \\
\hline
\phantom{0}1 & 000 & $(1-p_{i1})(1-p_{i2})(1-p_{i3})$ &....... & 1 & 100 \\
\phantom{0}2 & 100 & $p_{i1}\alpha(1-p_{i2})(1-p_{i3})$ & 1...... & 2 & 010 \\
\phantom{0}3 & 200 & $p_{i1}(1-\alpha)(1-p_{i2})(1-p_{i3})$ & 1...... & 3 & 110 \\
\phantom{0}4 & 010 & $(1-p_{i1})p_{i2}\alpha(1-p_{i3})$ &.2..... & 4 & 001 \\
\phantom{0}5 & 110 & $p_{i1}\alpha p_{i2}\alpha(1-p_{i3})$ &..3.... & 5 & 101 \\
\phantom{0}6 & 210 & $p_{i1}(1-\alpha)p_{i2}\alpha(1-p_{i3})$ & 12..... & 6 & 011
\\
\phantom{0}7 & 020 & $(1-p_{i1})p_{i2}(1-\alpha)(1-p_{i3})$ &.2..... & 7 & 111 \\
\phantom{0}8 & 120 & $p_{i1}\alpha p_{i2}(1-\alpha)(1-p_{i3})$ & 12..... & & \\
\phantom{0}9 & 220 & $p_{i1}(1-\alpha)p_{i2}(1-\alpha)(1-p_{i3})$ & 12..... & & \\
10 & 001 & $(1-p_{i1})(1-p_{i2})p_{i3}\alpha$ &...4... & & \\
11 & 101 & $p_{i1}\alpha(1-p_{i2})p_{i3}\alpha$ &....5.. & & \\
12 & 201 & $p_{i1}(1-\alpha)(1-p_{i2})p_{i3}\alpha$ & 1..4... & & \\
13 & 011 & $(1-p_{i1})p_{i2}\alpha p_{i3}\alpha$ &.....6. & & \\
14 & 111 & $p_{i1}\alpha p_{i2}\alpha p_{i3}\alpha$ &......7 & & \\
15 & 211 & $p_{i1}(1-\alpha)p_{i2}\alpha p_{i3}\alpha$ & 1....6. & & \\
16 & 021 & $(1-p_{i1})p_{i2}(1-\alpha)p_{i3}\alpha$ &.2.4... & & \\
17 & 121 & $p_{i1}\alpha p_{i2}(1-\alpha)p_{i3}\alpha$ &.2..5.. & & \\
18 & 221 & $p_{i1}(1-\alpha)p_{i2}(1-\alpha)p_{i3}\alpha$ & 12.4... & &
\\
19 & 002 & $(1-p_{i1})(1-p_{i2})p_{i3}(1-\alpha)$ &...4... & & \\
20 & 102 & $p_{i1}\alpha(1-p_{i2})p_{i3}(1-\alpha)$ & 1..4... & & \\
21 & 202 & $p_{i1}(1-\alpha)(1-p_{i2})p_{i3}(1-\alpha)$ & 1..4... & & \\
22 & 012 & $(1-p_{i1})p_{i2}\alpha p_{i3}(1-\alpha)$ &.2.4... & & \\
23 & 112 & $p_{i1}\alpha p_{i2}\alpha p_{i3}(1-\alpha)$ &..34... & & \\
24 & 212 & $p_{i1}(1-\alpha)p_{i2}\alpha p_{i3}(1-\alpha)$ & 12.4... &
& \\
25 & 022 & $(1-p_{i1})p_{i2}(1-\alpha)p_{i3}(1-\alpha)$ &.2.4... & & \\
26 & 122 & $p_{i1}\alpha p_{i2}(1-\alpha)p_{i3}(1-\alpha)$ & 12.4... &
& \\
27 & 222 & $p_{i1}(1-\alpha)p_{i2}(1-\alpha)p_{i3}(1-\alpha)$ & 12.4...
& & \\
\hline
\end{tabular*}\vspace*{-6pt}
\end{table}

We will for convenience refer to the latent and recorded histories
using indices. With three possible latent encounter types (0, 1 and 2),
the latent history for individual $i$, ${\mathbf h}_i=(h_{i1},h_{i2},\ldots,h_{iT})$, is identified by
\[
j=1+\sum_{t=1}^T h_{it}
3^{t-1},
\]
such that $H_i=j$ indicates individual $i$ has latent encounter history
$j$. For example, $H_i=16$ for $T=3$\vadjust{\goodbreak} indicates individual $i$ has
latent history ${\mathbf h}_i=021$, $H_i>1$ indicates individual $i$ was
encountered at least once, and Pr$(H_i=j)$ is the probability that
individual $i$ has latent history $j$. Similarly, a binary recorded
history ${\boldsymbol\omega}=(\omega_1,\omega_2,\ldots,\omega_T)$ is
identified by
\[
k=\sum_{t=1}^T \omega_t
2^{t-1},
\]
such that $f_k$ is the observed frequency of recorded history $k$.

To implement our method, it is necessary to construct a matrix ${\mathbf
A}$, such that $\mathbf{f}=\mathbf{A}^\prime\mathbf{x}$, where the latent history
frequency vector ${\mathbf x}$ has elements $x_j=\sum_i \mathrm{I}  (
H_i=j  )$ indicating the number of individuals with latent history
$j$, and $\mathrm{I}  ( H_i=j  )$ is an indicator function having
the value 1 when $H_i=j$ and 0 otherwise. The matrix $\mathbf{A}$ formally
describes the relationship between the recorded and latent histories,
and intuition about how $\mathbf{A}$ is constructed is best provided
through a simple example. Suppose $T=3$ for binary (i.e., detection,
nondetection) recorded histories as in Table~\ref{tabexample}. The $3^T
\times ( 2^T-1  )$ matrix $\mathbf{A}$ for this example can be
constructed from the corresponding contributed records column in
Table~\ref{tabexample} by simply replacing each dot ($.$) with a 0 and
any other entry with a 1.
Thus, the rows of ${\mathbf A}$ correspond to the $3^T$ possible latent
encounter histories and the columns correspond to the $2^T-1$ possible
recorded histories. For example, the sixth row of $\mathbf{A}$ indicates
that latent history 210 $(j=6)$ gives rise to the recorded histories
100 $(k=1)$ and 010 $(k=2)$ for binary recorded histories when $T=3$.

We treat the latent individual encounter histories as unobserved
quantities (just like $\bolds{\theta}$ and $\bolds{\rho}$) and use Bayesian
analysis methods to evaluate the joint posterior distribution
%
%
\begin{equation}
[ {\mathbf h},{\bolds\theta}, {\bolds\rho} \mid{\mathbf f} ] \propto [ {\mathbf h} \mid{\bolds
\theta},{\bolds\rho} ] [ {\mathbf f} \mid {\mathbf h},{\bolds\theta},{\bolds\rho} ] [ {\bolds\theta},
{\bolds\rho} ], \label{eqposterior}
\end{equation}
where $ [ {\mathbf f} \mid{\mathbf h},{\bolds\theta},{\bolds\rho}  ]=\mathrm{I}  ( \mathbf{A}^\prime\mathbf{x} = \mathbf{f}  )$. We note that $ [
{\mathbf f} \mid{\mathbf h},{\bolds\theta},{\bolds\rho}  ]$ does not depend on
${\bolds\theta}$ or ${\bolds\rho}$; the relation is deterministic rather
than stochastic in the cases we consider here. One of the keys to
sampling from equation (\ref{eqposterior}) using MCMC is proposing latent
history frequencies $\mathbf{x}$ that satisfy $\mathbf{A}^\prime\mathbf{x} = \mathbf
{f}$. This is accomplished by utilizing basis vectors for the null
space of $\mathbf{A}^\prime$. Once the $\mathbf{A}$ matrix is defined, a basis
for the null space of $\mathbf{A}^\prime$ can be determined by solving the
system of equations $\mathbf{A}^\prime\mathbf{x} = \mathbf{0}$. For binary
recorded histories with $T=2$, one such basis is the set of
$3^T-2^T+1=6$ column vectors $ \{ {\mathbf v}  \}$, where ${\mathbf
v}_1= ( 1,0,0,0,0,0,0,0,0  )^\prime$, ${\mathbf v}_2= (
0,-1,1,0,0,0,0,0,0  )^\prime$, ${\mathbf v}_3= (
0,-1,0,-1,0,1,0,0,0  )^\prime$, ${\mathbf v}_4= (
0,0,0,-1,0, 0,1,0,0  )^\prime$, ${\mathbf v}_5= (
0,-1,0, -1,0,\break 0,0,1,0  )^\prime$ and ${\mathbf v}_6= (
0,-1,0,-1,0,0,0,0,1  )^\prime$.

When there is no individual heterogeneity in parameters, one may
propose and update $\mathbf{x}$ from the set of basis vectors without
explicit consideration of $\mathbf{h}_i$ [\citet{LinkEtAl2010}].
However, when allowing for individual heterogeneity, one must
explicitly consider $\mathbf{h}_i$ for each individual in the population.
An efficient MCMC algorithm therefore needs to regularly propose
reasonable $\mathbf{h}_i$ in combinations that satisfy $\mathbf{A}^\prime\mathbf
{x} = \mathbf{f}$. As illustrated in Sections~\ref{subsubsecmodel1} and \ref
{subsubsecmodel2} for closed population abundance models, we accomplish
this by apportioning each latent history frequency $x_j$ to individuals
with probabilities proportional to $\mbox{Pr}  ( H_i = j  )$.

\subsubsection{Model \texorpdfstring{$M_{t,b,h,\alpha}$}{M{t,b,h,alpha}}}\label{subsubsecmodel1}
For illustration, we now focus our efforts on extending the closed
population capture--recapture model $M_{t,b,h}$ [\citet
{OtisEtAl1978,KingBrooks2008}], which estimates abundance ($N$)
assuming temporal variation, behavioral effects and individual
heterogeneity in detection probabilities. Our extension includes all of
these effects while accounting for misidentification; we denote this
model as $M_{t,b,h,\alpha}$. Before proceeding, we again note that our
proposed approach may be used for other capture--recapture models [e.g.,
\citet{WilliamsEtAl2002}] by modifying them accordingly for
misidentification; the mathematical form for the $M_{t,b,h,\alpha}$
likelihood is simply substituted directly for $[{\mathbf h} \mid\bolds{\theta
},{\bolds\rho}]$ in equation (\ref{eqposterior}).

We adopt a Bayesian perspective and utilize data augmentation both to
account for individuals that were never detected [e.g., \citet
{RoyleEtAl2007}] and to formulate a probit model for detection
probability [e.g., \citet{Albert-Chib1993}]. The data augmentation
framework is useful because of computational efficiencies it produces,
and our procedure treats $N$ as a binomial random variable with known
index $M$ (typically $M \gg N$) and parameter $\psi$. In this context,
$M$ is often described as a ``superpopulation'' size of indicators $q_i
\sim\operatorname{Bernoulli}(\psi)$, where individuals with $q_i=1$ are
considered ``real individuals'' or ``individuals available for
capture,'' and $N=\sum_{i=1}^M q_i$. For the $\sum_{j=2}^{3^T} x_j$
individuals with $H_i>1$, we know $q_i=1$. For the remaining $M-\sum_{j=2}^{3^T} x_j$ individuals that were never detected, $H_i=1$ and
$q_i$ is unknown. A closed population misidentification model allowing
temporal and\vadjust{\goodbreak} individual variation in detection probability may then be
represented as
\begin{eqnarray*}
q_i \mid\psi&\sim&\operatorname{Bernoulli} (\psi ),
\\
h_{it} \mid q_i,p_{it} &\sim&\operatorname{Categorical}
\bigl( 1-q_i p_{it},\alpha q_i
p_{it}, (1-\alpha) q_i p_{it} \bigr)
\end{eqnarray*}
for $h_{it} \in \{ 0,1,2  \}$, where $p_{it}$ is the
probability of detection for individual $i$ at time~$t$, and $\alpha$
is the probability that an individual is correctly identified, given
detection. Because we assume $N \mid\psi\sim\operatorname{Binomial}(M,\psi
)$, a judicious choice of prior can yield the desired prior for $N$
when marginalized over $\psi$. For example, $\psi\sim\operatorname{Beta}(1,1)$ produces a discrete uniform prior on $N$. 

As a more computationally efficient alternative to the ubiquitous logit
link function for heterogeneous detection probabilities in Bayesian
capture--recapture analyses [e.g., \citet
{Castledine1981,George-Robert1992,FienbergEtAl1999,RoyleEtAl2007,KingBrooks2008,Link2013}],
we use data augmentation to formulate a probit model, $p_{it}=\Phi
({\mathbf w}_{it}^\prime{\bolds\beta} + \gamma_i  )$, where $\Phi$ is
the standard normal cumulative distribution function, ${\mathbf w}_{it}$ is
a vector of covariates for individual $i$ at time $t$, ${\bolds\beta}$ is
a vector of regression coefficients, and $\gamma_i$ is an
individual-level effect. Let $y_{it}=\mathrm{I}  (h_{it}>0  )$ be
an indicator for the binary detection process, and let $\tilde{y}_{it}$
be a continuous latent version of this process, where $\tilde{y}_{it}
\mid{\bolds\beta},\gamma_i \sim{\mathcal N}  ( {\mathbf w}_{it}^\prime
{\bolds\beta} + \gamma_i,1  )$. Assuming $y_{it} = 1$ if $\tilde
{y}_{it}>0$ and $q_i=1$, and assuming $y_{it} = 0$ if $\tilde
{y}_{it}<0$ and $q_i=1$ or $q_i=0$, then it follows that $y_{it} \mid
q_i,{\tilde y}_{it} \sim\operatorname{Bernoulli}  ( q_i \mathrm{I}  (
{\tilde y}_{it}>0  )  )$. This approach shares some
similarities with recent extensions of the probit regression model of
\citet{Albert-Chib1993} to imperfectly-detected species occurrence
data [\citet{Dorazio-Rodriguez2012,JohnsonEtAl2013}], but our
extension allows for individual-level effects and a response variable
of unknown dimension.

For our probit model allowing temporal, behavioral and individual
effects in detection probability, we define ${\mathbf w}_{it} =  ( \mathrm{I}  (t=1 ),\mathrm{I}  (t=2 ),\ldots,\mathrm{I}
(t=T),\mathrm{I}  (t>C_i )  )$ and ${\bolds\beta} =
(\beta_1,\beta_2,\ldots,\beta_{T+1}  )$, where $C_i$ denotes the
first capture occasion for individual $i$ (with $C_i=\infty$ for
individuals with $H_i=1$). Given the recorded history frequencies $\mathbf
{f}=(f_1,f_2,\ldots,f_{2^T-1})$, the joint posterior distribution for
model $M_{t,b,h,\alpha}$ is then
%
%
\begin{eqnarray} \label{eqposterior1}
\bigl[ \mathbf{h},\mathbf{q},{\tilde{\mathbf y}},\bolds{\beta},\bolds{\gamma},\psi, \alpha,
\sigma_\gamma^2 \mid\mathbf{f} \bigr] & \propto& [ \mathbf{h} \mid\mathbf
{q},{\tilde{\mathbf y}},\alpha ] \mathrm{I} \bigl( \mathbf{A}^\prime\mathbf{x} = \mathbf{f}
\bigr)\nonumber
\\
& &{} \times [ \mathbf{q} \mid\psi ] [ {\tilde{\mathbf y}} \mid {\bolds\beta},{\bolds\gamma} ] [
\bolds{\beta} \mid{\bolds\mu}_\beta, {\bolds\Sigma}_\beta ] \bigl[ \bolds{
\gamma} \mid\sigma_\gamma^2 \bigr]
\\
& &{} \times [ \psi ] [ \alpha ] \bigl[ \sigma _\gamma^2
\bigr],\nonumber
\end{eqnarray}
where
\begin{eqnarray*}
[ \mathbf{h} \mid\mathbf{q},{\tilde{\mathbf y}},\alpha ] &\propto& \prod
_{i=1}^M \prod_{t=1}^T
\bigl\{ q_i \mathrm{I} ( {\tilde y}_{it} >0 ) \bigr
\}^{\mathrm{I} (h_{it}>0  )} \bigl\{ 1- q_i \mathrm{I} ({\tilde y}_{it} >0
) \bigr\} ^{\mathrm{I} (h_{it}=0
)}
\\
&&\hspace*{27pt}{}\times  \alpha^{\mathrm{I}  (h_{it}=1  )} (1-\alpha)^{\mathrm{I}
(h_{it}=2  )}
\end{eqnarray*}
%
and
\begin{eqnarray*}
&& \mathbf{x}= \Biggl( \sum_{i=1}^M \mathrm{I} (
H_i=1 ), \sum_{i=1}^M \mathrm{I}
( H_i=2 ), \ldots, \sum_{i=1}^M
\mathrm{I} \bigl( H_i=3^T \bigr) \Biggr) = (
x_1,x_2,\ldots, x_{3^T} ). \label{eqx}
\end{eqnarray*}

We complete our Bayesian formulation by assigning the priors
\begin{eqnarray*}
\bolds{\beta} \mid{\bolds\mu}_\beta, {\bolds\Sigma}_\beta& \sim& {
\mathcal N} ( {\bolds\mu}_\beta,{\bolds\Sigma}_\beta ),
\\
\gamma_i \mid\sigma_\gamma^2 & \sim& {\mathcal
N} \bigl( 0, \sigma _\gamma^2 \bigr),
\\
\alpha& \sim& \operatorname{Beta} ( a_\alpha,
b_\alpha ),
\\
\psi& \sim& \operatorname{Beta} ( a_\psi, b_\psi ),
\end{eqnarray*}
and\vspace*{1pt} $\sigma_\gamma^2 \sim\Gamma^{-1}  ( a_{\sigma_\gamma},
b_{\sigma_\alpha}  )$, where ${\bolds\mu}_\beta$ and ${\bolds\Sigma
}_\beta$ are the prior mean and covariance matrix for ${\bolds\beta}$. By
choosing $b_\psi=1$ and a very small positive value for $a_\psi$, one
can approximate the scale prior $ [ N  ] \propto1/N$
[\citet{Link2013}]. We note that simpler closed population
abundance models may be specified by modifying model $M_{t,b,h,\alpha}$
accordingly. For example, set $\beta_{T+1}=0$ to remove behavior
effects, set $\beta_1=\beta_2=\cdots=\beta_T$ to remove temporal
variation, or set $\gamma_i=0$ for $i=1,\ldots,M$ to remove individual effects.

Given the $3^T \times ( 2^T-1  )$ matrix $\mathbf{A}$ for binary
recorded histories and a set of basis vectors $ \{ \mathbf{v}  \}
= \{{\mathbf v}_1,{\mathbf v}_2,\ldots,{\mathbf v}_{3^T-2^T+1}  \}$ for
the null space of $\mathbf{A}^\prime$ (where ${\mathbf v}_1$ is the basis
vector corresponding to the all-zero latent history frequency), we
propose the following MCMC algorithm for sampling from the posterior
distribution of model $M_{t,b,h,\alpha}$ [equation (\ref{eqposterior1})].
We utilize Metropolis--Hastings updates for the latent encounter
histories, but our judicious choice of priors enables Gibbs updates for
${\mathbf q}$, $\tilde{{\mathbf y}}$ and all parameters:
\begin{enumerate}[10.]
\item[1.] Initialize all parameters and latent variables, including an
initial feasible set of $M$ latent individual histories $({\mathbf h})$
with corresponding frequencies ${\mathbf x}$ satisfying $\mathbf{A}^\prime\mathbf
{x} = \mathbf{f}$. One such initial vector ${\mathbf x}$ is readily available
by assuming $\alpha=1$, such that latent frequencies corresponding to
histories with 2's are zeros, with a one-to-one matching of the
remaining latent frequencies with the recorded history frequencies~$(\mathbf f)$. This creates $\sum_{k=1}^{2^T-1} f_k$ individual histories\vspace*{1pt}
(with corresponding $H_i>1$), none of which is the all-zero history. To
complete the initialization, assign $x_1=M-\sum_{k=1}^{2^T-1} f_k$
individuals to the all-zero history (with corresponding $H_i=1$).
\item[2.] Update $\tilde{y}_{it}$ for $i=1,\ldots,M$ and
$t=1,\ldots,T$ from the full conditional distribution:
\begin{eqnarray*}
\tilde{y}_{it} \mid\cdot\sim\cases{ {\mathcal TN}_{ ( 0, \infty )}
\bigl( {\mathbf w}_{it}^\prime {\bolds\beta} + \gamma_i,1
\bigr), &\quad if $h_{it}>0$ and $q_i=1$,
\vspace*{3pt}\cr
{\mathcal
TN}_{ (- \infty, 0  )} \bigl( {\mathbf w}_{it}^\prime {\bolds\beta} +
\gamma_i,1 \bigr), &\quad if $h_{it}=0$ and
$q_i=1$,
\vspace*{3pt}\cr
{\mathcal N} \bigl( {\mathbf w}_{it}^\prime{
\bolds\beta} + \gamma_i,1 \bigr), &\quad otherwise,}
\end{eqnarray*}
where ${\mathcal TN}_{ ( L, U  )}$ is a normal distribution
truncated at $L$ and $U$.
\item[3.] Update ${\bolds\beta}$ from the full conditional distribution:
\[
{\bolds\beta} \mid\cdot\sim{\mathcal N} \bigl( \bigl( {\bolds\Sigma
}_\beta^{-1} + {\mathbf W}^\prime{\mathbf W}
\bigr)^{-1} \bigl({\bolds\Sigma }_\beta^{-1} {\bolds
\mu}_\beta+ {\mathbf W}^\prime{ ( \tilde{\mathbf y} - {\bolds\gamma}\otimes{
\mathbf1}_T )} \bigr), \bigl( {\bolds\Sigma }_\beta^{-1} +
{\mathbf W}^\prime{\mathbf W} \bigr)^{-1} \bigr),
\]
where ${\mathbf W}$ is the $MT \times(T+1)$ design matrix with rows ${\mathbf
w}_{it}^\prime$ and ${\mathbf1}_T$ is the all-ones vector of length $T$.
\item[4.] Update $\gamma_i$ for $i=1,\ldots,M$ from the full conditional
distribution:
\[
\gamma_i \mid\cdot\sim{\mathcal N} \biggl( \frac{\sigma_\gamma^2 \sum_{t=1}^T  ( {\tilde y}_{it} - {\mathbf w}_{it}^\prime{\bolds\beta}
)}{1+T \sigma_\gamma^2},
\frac{\sigma_\gamma^2}{1+T \sigma_\gamma^2} \biggr).
\]
\item[5.] Update $\sigma_\gamma^2$ from the full conditional distribution:
\[
\sigma_\gamma^2 \mid\cdot\sim\Gamma^{-1} \biggl(
a_{\sigma_\gamma
}+\frac{M}{2}, b_{\sigma_\gamma}+\frac{{\bolds\gamma}^\prime{\bolds\gamma
}}{2} \biggr).
\]
\item[6.] Update $\alpha$ from the full conditional distribution:
\[
\alpha\mid\cdot\sim\operatorname{Beta} \Biggl( a_\alpha+ \sum
_{i=1}^M \sum_{t=1}^T
\mathrm{I} ( h_{it}=1 ), b_\alpha+ \sum
_{i=1}^M \sum_{t=1}^T
\mathrm{I} ( h_{it}=2 ) \Biggr).
\]
\item[7.] Update $q_i$ for the $x_1$ individuals with $H_i=1$
from the full conditional distribution by drawing from a Bernoulli
distribution with probability
\[
\operatorname{Pr} ( q_i =1 \mid H_i =1 ) =
\frac{\psi\prod_{t=1}^T  \{ 1-\Phi ({\mathbf w}_{it}^\prime{\bolds\beta} + \gamma
_i  )  \}}{\psi\prod_{t=1}^T  \{ 1-\Phi ({\mathbf
w}_{it}^\prime{\bolds\beta} + \gamma_i  )  \}+ ( 1-\psi
 )}.
\]
\item[8.] Update $\psi$ from the full conditional distribution:
\[
\psi\mid\cdot\sim\operatorname{Beta} \Biggl( a_\psi+ \sum
_{i=1}^M q_i, b_\psi+M-\sum
_{i=1}^M q_i \Biggr).
\]
\item[9.] Update the set of $M$ latent encounter histories
$ ( \mathbf{h}  )$ using a Metropolis--Hastings step.
\begin{longlist}[(a)]
\item[(a)] Set $H_i^*=H_i$ for $i=1,\ldots,M$ and
$x_1^*=x_1$. Randomly draw $r$ from the integer set $ \{ 2,\ldots,
3^T - 2^T + 1  \}$ corresponding to basis vectors $ \{ {\mathbf
v}_2,\ldots,{\mathbf v}_{3^T-2^T+1}  \}$. Next draw $k_r$ from a
discrete uniform distribution over the integers $ \{ -D_r,\ldots,
-1, 1,\ldots, D_r  \}$, where $D_r$ is a tuning parameter. Propose
a latent history frequency vector
\[
\mathbf{x}^*=\mathbf{x}+k_r \mathbf{v}_r.
\]
If\vspace*{1pt} any $x_j^*<0$ for $j=2,\ldots,3^T$ or $M-\sum_{j=2}^{3^T} x_j^* <
0$, go to step~10.

\item[(b)] Apportion ${\mathbf x}^*$ to individuals with probabilities
proportional to $\operatorname{Pr}  ( H_i = j  )$. With probability
0.5, continue to step~9(b)(i) followed by step 9(b)(ii);
otherwise proceed with step 9(b)(ii) followed by step~9(b)(i).
\begin{enumerate}[(ii)]
\item[(i)] For each $x_j^* < x_j$ $(j=2,\ldots,3^T)$, draw
a set $ \{ O_r^{j-}  \}= \{o_1^{j-},o_2^{j-},\ldots,\break  o_{k_r}^{j-}  \}$ of\vspace*{1pt} individuals (of size $k_r$) without
replacement from the $x_j$ individuals with capture history $j$ (i.e.,
$H_i^*=j$) with respective probabilities
\[
\operatorname{Pr} \bigl(H_i^{**} = 1 \bigr) = \prod
_{t=1}^T \bigl\{ 1-\Phi \bigl({\mathbf
w}_{it}^\prime{\bolds\beta} + \gamma_i \bigr) \bigr
\},
\]
and set $H_i^* = 1$ for individuals $i \in \{O_r^{j-}  \}$.
After cycling through each $j$ for which $x_j^* < x_j$ $(j=2,\ldots,3^T)$, set $x_1^* = \sum_{i=1}^M \mathrm{I}  ( H_i^* =1  )$.
\item[(ii)] For each $x_j^* > x_j$ $(j=2,\ldots,3^T)$, draw a
set $ \{ O_r^{j+}  \}= \{o_1^{j+},o_2^{j+},\ldots,\break  o_{k_r}^{j+}  \}$ of individuals\vspace*{1pt} (of size $k_r$) without
replacement from the $x_1^*$ individuals that were never captured with
respective probabilities
\begin{eqnarray*}
\operatorname{Pr} \bigl(H_i^{**} = j \bigr) & = & \prod
_{t=1}^T \Phi \bigl({\mathbf
w}_{it}^\prime{\bolds\beta} + \gamma_i
\bigr)^{\mathrm{I} (
h_{it}^{**} >0  )} \bigl\{ 1-\Phi \bigl({\mathbf w}_{it}^\prime{
\bolds \beta} + \gamma_i \bigr) \bigr\}^{\mathrm{I} ( h_{it}^{**} =0
)}
\\
&&\hspace*{14pt}{} \times\alpha^{\mathrm{I} ( h_{it}^{**} =1  )} ( 1-\alpha )^{\mathrm{I} ( h_{it}^{**} =2  )}.
\end{eqnarray*}
Set $H_i^* = j$ and $q_i^* = 1$ for individuals $i \in \{
O_r^{j+} \}$, and set $x_1^* = \sum_{i=1}^M \mathrm{I}  ( H_i^*
=1  )$. Cycle through each $j$ for which $x_j^* > x_j$ $(j=2,\ldots, 3^T)$.
\end{enumerate}
\item[(c)] Propose $q_i^*$ for the $x_1^*$ individuals with $H_i^*=1$ as in
step~7.
Accept the proposed latent histories (i.e., set $\mathbf{x}=\mathbf{x}^*$,
$H_i=H_i^*$ and $q_i=q_i^*$) with probability $\min ( 1,R_r
)$, where
\begin{eqnarray*}
R_r &=& \Biggl( \Biggl[ \prod_{i\dvtx q_i^*=1}  \Biggl\{ \sum_{j=1}^{3^T} \operatorname{Pr}  \bigl(H_i^{**}=j \bigr) \mathrm{I}  \bigl(H_i^*=j  \bigr)  \Biggr\}
 \Biggr]
 \\
&&\hspace*{39pt}{}\times   \bigl[ \mathbf{q}^* \mid\psi \bigr]  \bigl[ \mathbf{h} \mid\mathbf{h}^*,
\bolds{\beta},\bolds{\gamma}, \alpha \bigr]  [ \mathbf{q} \mid\psi, \bolds
{\beta},\bolds{\gamma}  ]\Biggr)
\\
&&{} \Big/
\Biggl( \Biggl[ \prod_{i\dvtx q_i=1}  \Biggl\{ \sum_{j=1}^{3^T} \operatorname{Pr}  \bigl(H_i^{**}=j \bigr) \mathrm{I}  (H_i=j
 )  \Biggr\}  \Biggr]
 \\
 &&\hspace*{45pt}{}\times [ \mathbf{q} \mid\psi ]  \bigl[ \mathbf
{h}^* \mid\mathbf{h}, \bolds{\beta}, \bolds{\gamma}, \alpha \bigr]  \bigl[ \mathbf
{q}^* \mid\psi, \bolds{\beta},\bolds{\gamma}  \bigr]\Biggr),
\end{eqnarray*}
$ [ \mathbf{h}^* \mid\mathbf{h}, \bolds{\beta}, \bolds{\gamma}, \alpha ]$
is the proposal density for $\mathbf{h}^*$, and $ [ \mathbf{q}^* \mid\psi,
\bolds{\beta},\bolds{\gamma}  ]$ is the proposal density for $\mathbf{q}^*$.
Here, $ [ \mathbf{h}^* \mid\mathbf{h}, \bolds{\beta},\bolds{\gamma}, \alpha
 ]$ is the product of the (ordered) conditional inclusion
probabilities, $\operatorname{Pr}(H_i^{**}=1)$ for $i \in \{O_r^{j-}
\}$ and $\operatorname{Pr}(H_i^{**}=j)$ for $i \in \{O_r^{j+} \}$,
that were, respectively, selected in\break steps~9(b)(i)  and 9(b)(ii)
under unequal probability sampling without replacement [e.g.,
\citet{Thompson1992}, page 53]:
\begin{eqnarray*}
&& \bigl[ \mathbf{h}^* \mid\mathbf{h}, \bolds{\beta},\bolds{\gamma}, \alpha \bigr]
\\
&&\qquad  =  \Biggl[
\mathop{\prod_{j\dvtx x_j^* < x_j;}}_{j>1} \prod
_{s=1}^{k_r} \frac
{\operatorname{Pr}  (H_{o_s^{j-}}^{**}=1 ) }{\sum_{i\dvtx H_i=j}
\operatorname{Pr}  ( H_i^{**}=1  ) - \sum_{m=1}^{s-1}
\operatorname{Pr}  ( H_{o_m^{j-}}^{**}=1  )} \Biggr]
\nonumber
\\
&&\quad\qquad{} \times \Biggl[ \mathop{\prod_{j\dvtx x_j^* > x_j;}}_{j>1}
\prod_{s=1}^{k_r} \frac{\operatorname{Pr}  (H_{o_s^{j+}}^{**}=j
)}{\sum_{i\dvtx H_i^*=1} \operatorname{Pr}  ( H_i^{**}=j  ) - \sum_{m=1}^{s-1} \operatorname{Pr}  ( H_{o_m^{j+}}^{**}=j  )} \Biggr]
\end{eqnarray*}
and
\begin{eqnarray*}
&& \bigl[ \mathbf{h} \mid\mathbf{h}^*, \bolds{\beta},\bolds{\gamma}, \alpha \bigr]
\\
&&\qquad  =  \Biggl[
\mathop{\prod_{j\dvtx x_j^* < x_j;}}_{j>1} \prod
_{s=1}^{k_r} \frac
{\operatorname{Pr}  (H_{o_s^{j-}}^{**}=j ) }{\sum_{i\dvtx H_i^*=1}
\operatorname{Pr}  ( H_i^{**}=j  ) - \sum_{m=1}^{s-1}
\operatorname{Pr}  ( H_{o_m^{j-}}^{**}=j  )} \Biggr]
\nonumber
\\
&&\quad\qquad{} \times \Biggl[ \mathop{\prod_{j\dvtx x_j^* > x_j;}}_{j>1}
\prod_{s=1}^{k_r} \frac{\operatorname{Pr}  (H_{o_s^{j+}}^{**}=1 )
}{\sum_{i\dvtx H_i^*=j} \operatorname{Pr}  ( H_i^{**}=1  ) - \sum_{m=1}^{s-1} \operatorname{Pr}
( H_{o^{j+}_m}^{**}=1  )} \Biggr].
\end{eqnarray*}
%
\end{longlist}
\item[10.] Return to step 2 and repeat as needed.
\end{enumerate}
Note that $N$ is obtained by calculating $N=\sum_{i=1}^M q_i$ at each
iteration of the algorithm.

\subsubsection{Model \texorpdfstring{$M_{t,b,\alpha_h}$}{M{t,b,alphah}}}\label{subsubsecmodel2}
In some applications, one may be more concerned about individual
heterogeneity in misidentification than detection probability. For
example, the quality of visual identifiers (e.g., artificial marks,
naturally occurring pelt or scar patterns) or genetic material (e.g.,
hair or fecal samples) may vary by individual [\citet
{LukacsBurnham2005}], and some individuals may therefore be more or
less likely to spawn ghost histories (see \textit{Blue Ridge
two-lined salamander}). We can modify model $M_{t,b}$ [\citet
{OtisEtAl1978,KingBrooks2008}] to accommodate temporal variation,
behavioral effects and individual heterogeneity in correct
identification probability, obtaining a model we call $M_{t,b,\alpha_h}$.

Similar to Section~\ref{subsubsecmodel1}, we specify a probit model for
the probability of correctly identifying an individual, given
detection, $\alpha_i=\Phi ( \mu_\alpha+ \varepsilon_i  )$,
where $\mu_\alpha$ is an intercept term and $\varepsilon_i$ is an
individual-level effect. Let $u_{it}$ be an indicator for the binary
correct identification process, and let $\tilde{u}_{it}$ be a
continuous latent version of this process, where $\tilde{u}_{it} \mid
\mu_\alpha, \varepsilon_i \sim{\mathcal N}  ( \mu_\alpha+ \varepsilon
_i,1  )$. Assuming $u_{it} = 1$ if $\tilde{u}_{it}>0$ and
$h_{it}=1$, and assuming $u_{it} = 0$ if $\tilde{u}_{it}<0$ and
$h_{it}=2$ or $h_{it}=0$, then it follows that\vadjust{\goodbreak} $u_{it} \mid
h_{it},{\tilde u}_{it} \sim\operatorname{Bernoulli}  ( \mathrm{I}  (
h_{it}>0  ) \mathrm{I}  ( {\tilde u}_{it}>0  )  )$. The
joint posterior distribution for model $M_{t,b,\alpha_h}$ is then
%
%
\begin{eqnarray} \label{eqposterior2}
\bigl[ \mathbf{h},\mathbf{q},\tilde{{\mathbf y}},\tilde{{\mathbf u}},\bolds{\beta},\mu
_\alpha,\bolds{\varepsilon},\psi,\sigma_\varepsilon^2 \mid
\mathbf{f} \bigr] & \propto& [ \mathbf{h} \mid\mathbf{q},{\tilde{\mathbf y}},{\tilde{\mathbf u}} ] \mathrm{I} \bigl( \mathbf{A}^\prime\mathbf{x} = \mathbf{f} \bigr)
\nonumber
\\
& &{} \times [ \mathbf{q} \mid\psi ] [ {\tilde{\mathbf y}} \mid {\bolds\beta} ] [ {\tilde{\mathbf
u}} \mid\mu_\alpha,{\bolds \varepsilon} ] [ \bolds{\beta} \mid{\bolds
\mu}_\beta, {\bolds\Sigma }_\beta ] \bigl[ \bolds{\varepsilon} \mid
\sigma_\varepsilon^2 \bigr]
\\
& &{} \times [ \psi ] [ \mu_\alpha ] \bigl[ \sigma _\varepsilon^2
\bigr],\nonumber
\end{eqnarray}
where
\begin{eqnarray*}
[ \mathbf{h} \mid\mathbf{q},{\tilde{\mathbf y}},{\tilde{\mathbf u}} ] & \propto& \prod
_{i=1}^M \prod_{t=1}^T
\bigl\{ q_i \mathrm{I} ( {\tilde y}_{it} >0 ) \bigr
\}^{\mathrm{I} (h_{it}>0  )} \bigl\{ 1- q_i \mathrm{I} ({\tilde y}_{it} >0
) \bigr\} ^{\mathrm{I} ( h_{it}=0  )}
\\
&&\hspace*{27pt} {} \times \bigl\{ \mathrm{I} ( h_{it}>0 ) \mathrm{I} ( {\tilde
u}_{it} >0 ) \bigr\} ^{\mathrm{I}  (h_{it}=1  )}
\\
&&\hspace*{27pt} {} \times
\bigl\{ 1-\mathrm{I} (
h_{it}>0 ) \mathrm{I} ({\tilde u}_{it} >0 ) \bigr
\}^{\mathrm{I}
(h_{it}=2  )},
\end{eqnarray*}
${\tilde y}_{it} \mid{\bolds\beta} \sim{\mathcal N}  ( {\mathbf
w}_{it}^\prime{\bolds\beta},1  )$, and all other components of the
model are specified as in Section~\ref{subsubsecmodel1} for model
$M_{t,b,h,\alpha}$. We assign the additional priors $\mu_\alpha\sim
{\mathcal N}  ( \mu_{\mu_\alpha}, \sigma^2_{\mu_\alpha}  )$,
$\varepsilon_i \mid\sigma_\varepsilon^2 \sim{\mathcal N}  ( 0, \sigma
_\varepsilon^2  )$, and $\sigma_\varepsilon^2 \sim\Gamma^{-1}  (
a_{\sigma_\varepsilon}, b_{\sigma_\varepsilon}  )$. 

It is straightforward to modify the MCMC algorithm described in
Section~\ref{subsubsecmodel1} for sampling from the posterior
distribution of model $M_{t,b,\alpha_h}$ [equation (\ref{eqposterior2})].
The additional parameters and ${\tilde{\mathbf u}}$ are simply updated
from their full conditional distributions:
\begin{eqnarray*}
{\tilde u}_{it} \mid\cdot&\sim& \cases{ {\mathcal TN}_{ ( 0, \infty )} (
\mu_\alpha+ \varepsilon _i,1 ), &\quad if $h_{it}=1$,
\vspace*{3pt}\cr
{\mathcal TN}_{ (- \infty, 0  )} ( \mu_\alpha+ \varepsilon _i,1 ),&\quad if $h_{it}=2$,
\vspace*{3pt}\cr
{\mathcal N} ( \mu_\alpha+
\varepsilon_i,1 ), &\quad otherwise,}
\\
\mu_\alpha\mid\cdot&\sim&{\mathcal N} \Biggl( \biggl( \frac{1}{\sigma
_{\mu_\alpha}^2}+
MT \biggr)^{-1} \Biggl( \frac{\mu_{\mu_\alpha}}{\sigma
_{\mu_\alpha}^2} + \sum
_{i=1}^M \sum_{t=1}^T
\{ {\tilde u}_{it}-\varepsilon_i \} \Biggr), \biggl(
\frac{1}{\sigma_{\mu_\alpha
}^2}+ MT \biggr)^{-1} \Biggr),
\\
\varepsilon_i \mid\cdot&\sim&{\mathcal N} \biggl( \frac{\sigma_\varepsilon^2
\sum_{t=1}^T  \{ {\tilde u}_{it}-\mu_\alpha \} }{1+T\sigma
_\varepsilon^2},
\frac{\sigma_\varepsilon^2}{1+T\sigma_\varepsilon^2} \biggr)
\end{eqnarray*}
and
\[
\sigma_\varepsilon^2 \mid\cdot\sim\Gamma^{-1} \biggl(
a_{\sigma_\varepsilon
}+\frac{M}{2},b_{\sigma_\varepsilon}+\frac{{\bolds\varepsilon}^\prime{\bolds
\varepsilon}}{2} \biggr).
\]
The only other notable difference from our algorithm for model
$M_{t,b,h,\alpha}$ is that we instead use $\operatorname{Pr}  ( H_i^{**}=1
 ) = 1 - \operatorname{Pr}  (H_i^{**} = j  )$ to propose
individuals that were never detected in the step corresponding to 9(b)(i)
above. This is because under model $M_{t,b,\alpha_h}$, all
individuals have the same probability of never being detected.

\section{Example applications}\label{secexamples}

\subsection{Black bears of the Northern Lower Peninsula, Michigan, USA}
\label{subsecbears}
In an impressive field and analytical effort, \citet
{DreherEtAl2007} applied a closed\vadjust{\goodbreak} population model that incorporates
individual misidentification due to genotyping error [\citet
{LukacsBurnham2005}] to estimate black bear (\textit{Ursus americanus})
abundance in the Northern Lower Peninsula of Michigan, USA. DNA samples
were collected from baited barbed wire hair snares on five occasions
from 22 June--26~July 2003. A sixth DNA sampling occasion occurred
through the extraction of teeth and muscle tissue from bears registered
during the recreational harvest in the autumn (hence, $T=6$). In
addition, a random sample of hand-pulled hair samples collected from
harvested bears provided auxiliary information about the probability of
a genotyping error using hair-snare samples. Complete details of the
data collection, genetic analysis and statistical analysis can be found
in \citet{DreherEtAl2007}.

Here we re-visit the DNA capture--recapture data of \citet
{DreherEtAl2007} using our closed population abundance model allowing
for temporal variation, behavioral effects, individual heterogeneity
and misidentification (Section~\ref{subsubsecmodel1}). Our motivation
is twofold: (1) individual heterogeneity in detection from hair-snare
samples was suspected by \citet{DreherEtAl2007}, but not
incorporated into their misidentification model; and (2) the
misidentification model proposed by \citet{LukacsBurnham2005}
relies on several assumptions that are unlikely to be met in practice
and does not properly account for ghost capture histories that result
from misidentification [\citet
{Yoshizaki2007,LinkEtAl2010,YoshizakiEtAl2011}].

Based on the best-supported model from \citet{DreherEtAl2007},
we fit model $M_{\hunt,b,h,\alpha}$, which allows for different
detection probabilities for the two methods of capture (i.e., hair
snare or harvest; indicated by ``hunt''), a behavioral response to the
baited hair snares, individual heterogeneity in detection probability
from hair-snare sampling, and misidentification of hair snare samples
due to genotyping error. Allowing misidentification to occur only for
the hair-snare sampling occasions $(t=1,\ldots,5)$, we have
\begin{eqnarray*}
[ \mathbf{h} \mid\mathbf{q},{\tilde{\mathbf y}},\alpha ] & \propto& \prod
_{i=1}^M \Biggl[ \prod_{t=1}^5
\bigl\{ q_i \mathrm{I} ( {\tilde y}_{it} >0 ) \bigr
\}^{\mathrm{I} (h_{it}>0  )} \bigl\{ 1- q_i \mathrm{I} ({\tilde y}_{it} >0
) \bigr\} ^{\mathrm{I}
(h_{it}=0  )}
\\
&&\hspace*{121pt}{}\times \alpha^{\mathrm{I}  (h_{it}=1  )} (1-\alpha )^{\mathrm{I}  (h_{it}=2  )}
\Biggr]
\nonumber
\\
&&\hspace*{12pt}{} \times ( q_i p_{\hunt} )^{h_{i6}} (
1-q_i p_{\hunt} )^{1-h_{i6}},
\end{eqnarray*}
$p_{it}=\Phi ( {\mathbf w}_{it}^\prime{\bolds\beta} +\gamma_i  )$
for $t=1,\ldots,5$, ${\mathbf w}_{it}= (1, \mathrm{I}  (t>C_i  )
 )$, ${\bolds\beta} =  ( \beta_1, \beta_2  )$, and
$p_{\hunt} \mid a_p,b_p \sim\operatorname{Beta}  ( a_p,b_p  )$.

The ${\mathbf A}$ matrix, posterior and MCMC algorithm described in
Section~\ref{subsubsecmodel1} are modified accordingly, where the
reduced $ [ 2  ( 3^{T-1} )  ] \times (2^T-1
)$ ${\mathbf A}$ matrix does not include misidentification for the harvest
sampling occasion $(t=6)$, and $p_{\hunt}$ is updated from the full
conditional distribution: $p_{\hunt} \mid\cdot\sim\operatorname{Beta}  (
a_p+\sum_{i=1}^M q_i h_{i6},b_p+\sum_{i=1}^M q_i  ( 1- h_{i6}
)  )$. We used weakly informative priors by setting ${\bolds\mu
}_\beta= {\mathbf0}$ and ${\bolds\Sigma}_\beta= \operatorname{diag}  ( 10,10
 )$, $a_\psi=10^{-6}$ and $a_{\sigma_\gamma}=b_{\sigma_\gamma
}=a_p=b_p=b_\psi=1$. Based on the auxiliary data about genotyping error
of hair samples collected from harvested bears (where 91 out of 95
samples were correctly assigned to genotype), we used an informative
prior with $a_\alpha=91$ and $b_\alpha=4$. To investigate prior
sensitivity, we conducted an additional analysis using an uninformative
prior on $\alpha$ by specifying $a_\alpha=b_\alpha=1$.

%
\begin{figure}

\includegraphics{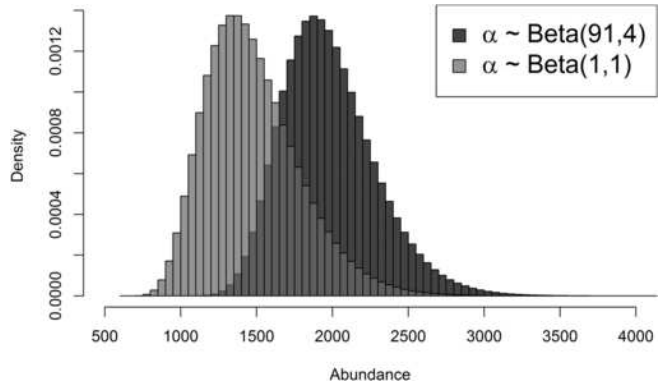}

\caption{Posterior distributions for abundance of black bears in the
Northern Lower Peninsula of Michigan, USA, from DNA capture--recapture
surveys conducted in summer and autumn 2003. Results are for analyses
using an uninformative prior (light) and an informative prior (dark) on
the probability of correctly identifying an individual, given detection
$(\alpha)$.}
\label{figbearabundance}
\end{figure}

Our MCMC algorithm was written in the C programming language\break 
[\citet{KernighanRitchie1988}] with data pre- and post-processing
performed in R via the .C interface [\citet{RTeam2012}]. Starting
with $D_r=1$ and rounding to the nearest integer, we tuned the MH
sampler every 5000 iterations by multiplying or dividing $D_r$ by 0.95
if the acceptance rate for basis vector $r$ was $\le$0.44 or $>$0.44,
respectively, where acceptance rates were calculated as the number of
accepted moves divided by the number of attempted moves. After pilot
tuning and burn-in of 500,000 iterations from overdispersed starting
values, we obtained three chains of 10 million iterations for both
analyses. With $M=5000$, our analyses required about 48~hrs on a
computer running 64-bit Windows 7 (3.4~GHz Intel Core i7 processor, 16~Gb
RAM). Slow mixing necessitated long runs, likely due to correlated
parameters and low movement rates for the MH sampler. Similar to
\citet{LinkEtAl2010}, low movement rates for the MH sampler
resulted from many of the 486 possible latent histories having very low
probability. Chain convergence was assessed by visual inspection and
the Gelman--Rubin--Brooks (GRB) diagnostic in the R package CODA
[\citet{PlummerEtAl2006}]. For both analyses, all univariate GRB
diagnostics were $<$1.002 and the multivariate GRB diagnostic was 1.001
for monitored parameters $(N,\beta_1,\beta_2,\sigma_\gamma
^2,p_{\hunt},\alpha)$. Based on sample autocorrelations, mixing was
somewhat slower using the uninformative prior on $\alpha$, but
effective sample sizes exceeded 5000 for all parameters.

Using the informative prior on $\alpha$, we estimated posterior median
$N=1945$ with a 95\% credible interval (CI) of 1470--2681 (Figure~\ref
{figbearabundance}). Similar to \citet{DreherEtAl2007}, we found
evidence of a trap ``happy'' behavioral response to the baited hair
snares, with posterior mean $\beta_2=0.50$ $(95\% \mbox{
CI}={}$0.08--0.89). Estimates for $p_{\hunt}$ suggest about 21\% of this
population was harvested and reported to officials (Table~\ref{tabbears}).
%

%
%
\begin{table}
\tabcolsep=0pt
\caption{Posterior summaries and effective sample sizes (ESS) for model
$M_{\hunt,b,h,\alpha}$ using black bear DNA capture--recapture data
collected in the Northern Lower Peninsula of Michigan, USA in 2003.
Mean capture and recapture probabilities were derived as $\bar{p}=\int_{-\infty}^\infty\Phi ( \beta_1 + \gamma )  [ \gamma
\mid\sigma_\gamma^2  ] \,d \gamma$ and $\bar{c}=\int_{-\infty
}^\infty\Phi ( \beta_1 + \beta_2 + \gamma )  [ \gamma
\mid\sigma_\gamma^2  ] \,d \gamma$, respectively}\label{tabbears}
\begin{tabular*}{\tablewidth}{@{\extracolsep{\fill}}@{}ld{4.2}d{4.2}d{4.2}d{4.2}d{4.2}d{4.2}c@{}}
\hline
& & & & & \multicolumn{2}{c}{\textbf{95\%}}  \\[-6pt]
& & & & & \multicolumn{2}{c}{\hrulefill}\\
\textbf{Parm.} & \multicolumn{1}{c}{\textbf{Mean}} & \multicolumn{1}{c}{\textbf{Median}} & \multicolumn{1}{c}{\textbf{Mode}}
& \multicolumn{1}{c}{\textbf{SD}} & \multicolumn{1}{c}{\textbf{LCI}} & \multicolumn{1}{c}{\textbf{UCI}} & \textbf{ESS}\\
\hline
$N$ & 1978.9 & 1945 & 1875 & 310.5 & 1470 & 2681 & 69,711 \\
$\bar{p}$ & 0.02 & 0.02 & 0.02 & 0.00 & 0.01 & 0.03 & 89,171 \\
$\bar{c}$ & 0.05 & 0.05 & 0.04 & 0.02 & 0.02 & 0.10 & 31,380 \\
$p_{\hunt}$ & 0.21 & 0.21 & 0.21 & 0.03 & 0.15 & 0.28 & 80,241 \\
$\alpha$ & 0.95 & 0.96 & 0.96 & 0.02 & 0.90 & 0.99 & 11,107 \\
$\beta_1$ & -2.48 & -2.46 & -2.44 & 0.16 & -2.84 & -2.21 & 16,568 \\
$\beta_2$ & 0.50 & 0.51 & 0.53 & 0.21 & 0.08 & 0.89 & 55,550 \\
$\sigma_\gamma$ & 0.64 & 0.63 & 0.61 & 0.13 & 0.42 & 0.92 & 16,371 \\
\hline
\end{tabular*}
\end{table}
As suspected by \citet{DreherEtAl2007}, we found evidence of
individual heterogeneity in detection from hair snares, with posterior
median $\sigma_\gamma=0.63$ $(95\% \mbox{ CI}={}$0.42--0.92). Because
unmodeled individual heterogeneity tends to cause underestimation of
abundance, this likely explains our posterior distribution for $N$
having support at higher values than the original estimates using
models that did not account for individual heterogeneity. For example,
\citet{DreherEtAl2007} estimated $N=1882$ with a 95\% confidence
interval of 1389--2551 for model $M_{\hunt,b}$ using the
misidentification model proposed by \citet{LukacsBurnham2005}.

%
%
\begin{figure}

\includegraphics{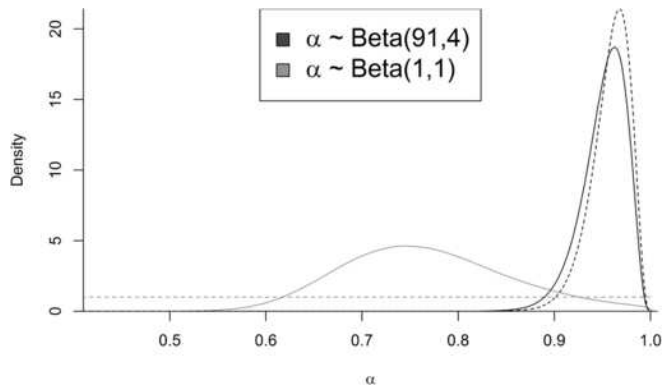}

\caption{Posterior (solid lines) and prior (dashed lines) densities for
the probability of correctly genotyping DNA hair-snare samples $(\alpha
)$ collected from black bears in the Northern Lower Peninsula of
Michigan, USA. Results are for analyses using an uninformative prior
(light) and an informative prior (dark) on $\alpha$.}
\label{figbearalpha}
\end{figure}

We estimated posterior mean $\alpha=0.95$ $(95\% \mbox{
CI}={}$0.90--0.99), with slight evidence of higher misidentification
probabilities from the hair-snare samples than\vadjust{\goodbreak} from the auxiliary hair
samples collected from harvested bears (Figure~\ref{figbearalpha}). The
auxiliary genotyping error data proved quite informative; the analogous
analysis using an uninformative $\operatorname{Beta} (1,1)$ prior on $\alpha$
yielded posterior median $N=1436$ ($95\% \mbox{ CI}={}$988--2203;
Figure~\ref{figbearabundance}) and posterior mean $\alpha=0.77$ ($95\%
\mbox{ CI}={}$0.61--0.95; Figure~\ref{figbearalpha}). In the absence of
prior information, the recorded histories may only provide minimal
information about misidentification, such as the range of $\alpha$ for
which there is very little support. Put another way, the frequencies of
potential ghost histories alone suggest $\alpha>0.55$. If $\alpha$ were
in fact $\ll$0.55, we would expect many more ghost histories to have
been observed relative to the observed nonghost histories [see
Section~4.1 in \citet{LinkEtAl2010} for further discussion].
Nevertheless, when using the informative prior for $\alpha$, we found
relatively little contrary information about misidentification from the
recorded histories. Given this prior sensitivity, care should be taken
in specifying informative priors for $\alpha$. For example, there could
be reason to suspect that hair samples collected from harvested bears
are of higher quality than hair-snare samples (e.g., due to degradation
by environmental factors), in which case hair-snare misidentification
could potentially be underestimated (and abundance overestimated) from
this prior.

\subsection{Blue Ridge two-lined salamanders}\label{subsecsalamanders}
Bailey (\citeyear{Bailey2004}) conducted a laboratory experiment evaluating
the ability of observers to individually identify Blue Ridge two-lined
salamanders (\textit{Eurycea bislineata wilderae}) marked with a
subcutaneous injection of elastomer (a silicone-based material
manufactured by Northwest Marine Technology, Inc., Shaw Island,
Washington, USA). Out of a pool of 20 marked salamanders, each of 14
observers viewed 10 randomly chosen individuals. Two different lights
were used for viewing the marks: a dive light with blue filter lens
(hereafter blue light) and a deep blue 7-LED flashlight (hereafter
black light). Observers first viewed each salamander with one light
(randomly assigned), and then the 10 individuals were re-randomized and
presented to the observers for identification using the other light.
\citet{Bailey2004} found no difference in observer ability to
correctly identify individuals based on the light used, but found mark
quality strongly influenced observers' ability to correctly identify
individuals. For example, one individual salamander accounted for 10 of
15 misidentifications resulting from missed marks (because one of its
marks was quite small).

This laboratory experiment affords an opportunity to apply model
$M_{t,\alpha_h}$ on a population of known size $(N=20)$ with no
individual variation in detection and suspected individual variation in
misidentification probability. Although some ghosts were identified by
multiple observers, here we analyze a subset of $T=8$ observers for
which all ghost encounter histories contain a single detection and no
identification errors matched a legitimate individual (thus satisfying
these assumptions of the model). Because the true encounter history for
each marked individual was known, recorded history data were simply
generated from the true histories. For example, suppose an individual
was presented to 4 of the 8 observers and, using the blue light, they
recorded the true encounter history .12...21 (where a dot indicates
this individual was not presented to the corresponding observer), then
the blue light recorded histories spawned from this true encounter
history would be 01000001, 00100000 and 00000010. We performed separate
analyses for the blue and black light recorded histories to examine
potential differences in misidentification probabilities, as well as
our model's ability to accurately estimate the number of salamanders
used in the experiment.

Allowing for temporal variation in detection and individual variation
in misidentification probability, we modify the posterior and MCMC
algorithm described in Section~\ref{subsubsecmodel2} accordingly.
Setting $p_{it}=p_t$ and assuming $p_t \mid a_p, b_p \sim\operatorname{Beta}
 (a_p,b_p )$, $p_t$ can be updated from the full conditional
distribution: $p_t \mid\cdot\sim\operatorname{Beta}  ( a_p + \sum_{i=1}^M q_i \mathrm{I}  ( h_{it}>0  ), b_p + \sum_{i=1}^M q_i
\mathrm{I}  ( h_{it}=0  )  )$. We used weakly informative
priors by setting $\mu_{\mu_\alpha}=0$, $\sigma_{\mu_\alpha}^2=10$,
$a_\psi=10^{-6}$, and $a_{\sigma_\varepsilon}=b_{\sigma_\varepsilon
}=a_p=b_p=b_\psi=1$.

For both the blue and black light analyses, we obtained three chains of
10 million iterations after initial pilot tuning and a burn-in of
500,000 iterations from overdispersed starting values. With $M=200$, our
analyses required about 2~hrs to complete. As in the black bear
example, relatively slow mixing necessitated long runs, likely due to
correlated parameters and low movement rates for the MH sampler. For
both analyses, all univariate GRB diagnostics were $<$1.05 and the
multivariate GRB diagnostic was $<$1.008 for monitored parameters
$(N,p_t,\mu_\alpha,\sigma_\varepsilon^2,\alpha)$. Based on sample
autocorrelations, mixing was somewhat slower for several parameters in
the blue light analyses, but all effective sample sizes exceeded 8000
for both analyses.

For the blue light analysis, we found posterior median $N=20$ with a
95\% credible interval of 18--23. For the black light analysis, we
found posterior median $N=21$ $(95\% \mbox{ CI}={}$19--25) (Table~\ref
{tabsalamanders}). Hence, our model was able to reliably estimate $N$
using either light source. As in \citet{Bailey2004}, we found
misidentification probabilities were similar for the blue and black
lights, with posterior mean $\bar{\alpha}=\int_{-\infty}^\infty\Phi
 ( \mu_\alpha+ \varepsilon )  [ \varepsilon\mid\sigma
_\varepsilon^2  ] \,d \varepsilon= 0.88$ $(95\% \mbox{ CI}={}$0.74--0.97)
and 0.88 $(95\% \mbox{ CI}={}$0.76--0.97), respectively. We found some
evidence of individual heterogeneity in misidentification probabilities
attributable to variable mark quality, with posterior median $\sigma
_\varepsilon=1.34$ $(95\% \mbox{ CI}={}$0.54--4.51) and $\sigma_\varepsilon
=1.07$ $(95\% \mbox{ CI}={}$0.50--3.45) for the blue and black lights,
respectively.
%
%
\begin{sidewaystable}
\tablewidth=\textwidth
\tabcolsep=0pt
\caption{Posterior summaries and effective sample sizes (ESS) for model
$M_{t,\alpha_h}$ using salamander data generated from laboratory
experiments evaluating the effectiveness of a subcutaneously injected
marking material with two light sources (blue and black). The
population was of known size $(N=20)$ with 8 total misidentifications
(0--3 per individual) using the blue light and 9 total
misidentifications (0--2 per individual) using the black light}\label{tabsalamanders}
\begin{tabular*}{\tablewidth}{@{\extracolsep{\fill}}ld{2.2}d{2.2}d{2.2}d{1.2}d{2.2}d{2.2}d{6.0}
d{2.2}d{2.2}d{2.2}d{1.2}d{2.2}d{2.2}d{6.0}@{\hspace*{-1pt}}}
\hline
& \multicolumn{7}{c}{\textbf{Blue light}} & \multicolumn{7}{c@{}}{\textbf{Black light}}\\[-6pt]
& \multicolumn{7}{c@{\hspace*{2pt}}}{\hrulefill} &  \multicolumn{7}{c@{}}{\hrulefill}\\
 & & & & \multicolumn{2}{c}{\textbf{95\%}} & & & & &  & \multicolumn{2}{c}{\textbf{95\%}}\\[-6pt]
 & & & & \multicolumn{2}{c}{\hrulefill} & & & & & &  \multicolumn{2}{c}{\hrulefill}\\
\textbf{Parm.} & \multicolumn{1}{c}{\textbf{Mean}} & \multicolumn{1}{c}{\textbf{Median}}
               & \multicolumn{1}{c}{\textbf{Mode}} & \multicolumn{1}{c}{\textbf{SD}}
               & \multicolumn{1}{c}{\textbf{LCI}} & \multicolumn{1}{c}{\textbf{UCI}}
               & \multicolumn{1}{c}{\textbf{ESS}}  &  \multicolumn{1}{c}{\textbf{Mean}}
               & \multicolumn{1}{c}{\textbf{Median}} & \multicolumn{1}{c}{\textbf{Mode}}
               & \multicolumn{1}{c}{\textbf{SD}} & \multicolumn{1}{c}{\textbf{LCI}}
               & \multicolumn{1}{c}{\textbf{UCI}} & \multicolumn{1}{c@{}}{\textbf{ESS}}\\
\hline
$N$ & 19.8 & 20 & 20 & 1.3 & 18 & 23 & 11{,}935 &  21.3 & 21 & 21 & 1.7 &19 & 25 & 34{,}562 \\
$p_1$ & 0.51 & 0.51 & 0.51 & 0.11 & 0.30 & 0.72 & 1{,}378{,}542 &  0.47 &0.47 & 0.46 & 0.11 & 0.27 & 0.68 & 924{,}064 \\
$p_2$ & 0.51 & 0.51 & 0.50 & 0.11 & 0.30 & 0.72 & 1{,}383{,}346 &  0.47 &0.47 & 0.47 & 0.11 & 0.27 & 0.68 & 925{,}735 \\
$p_3$ & 0.51 & 0.51 & 0.51 & 0.11 & 0.30 & 0.72 & 1{,}379{,}963 &  0.47 &0.47 & 0.47 & 0.11 & 0.27 & 0.68 & 922{,}474 \\
$p_4$ & 0.51 & 0.51 & 0.51 & 0.11 & 0.30 & 0.72 & 1{,}381{,}844 &  0.43 &0.43 & 0.42 & 0.10 & 0.24 & 0.64 & 1{,}175{,}373 \\
$p_5$ & 0.51 & 0.51 & 0.51 & 0.11 & 0.30 & 0.72 & 1{,}389{,}232 &  0.47 &0.47 & 0.47 & 0.11 & 0.27 & 0.68 & 924{,}540 \\
$p_6$ & 0.51 & 0.51 & 0.50 & 0.11 & 0.30 & 0.72 & 1{,}379{,}287 &  0.47 &0.47 & 0.47 & 0.11 & 0.27 & 0.68 & 924{,}655 \\
$p_7$ & 0.51 & 0.51 & 0.51 & 0.11 & 0.30 & 0.72 & 1{,}384{,}098 &  0.47 &0.47 & 0.47 & 0.11 & 0.27 & 0.68 & 920{,}696 \\
$p_8$ & 0.51 & 0.51 & 0.50 & 0.11 & 0.30 & 0.72 & 1{,}381{,}843 &  0.47 &0.47 & 0.47 & 0.11 & 0.27 & 0.68 & 927{,}481 \\
$\mu_\alpha$ & 2.37 & 2.06 & 1.62 & 1.11 & 1.05 & 5.37 & 8667 &  2.06& 1.83 & 1.59 & 0.88 & 1.02 & 4.44 & 11{,}134 \\
$\sigma_\varepsilon$ & 1.65 & 1.34 & 0.88 & 1.06 & 0.54 & 4.51 & 8264 & 1.30 & 1.07 & 0.79 & 0.80 & 0.50 & 3.45 & 10{,}593 \\
$\alpha$ & 0.88 & 0.89 & 0.90 & 0.06 & 0.74 & 0.97 & 58{,}385 &  0.88 &0.89 & 0.90 & 0.05 & 0.76 & 0.97 & 71{,}694 \\
\hline
\end{tabular*}
\end{sidewaystable}

\section{Discussion}\label{secdiscussion}
We have presented a general model formulation and\break MCMC model-fitting
algorithm for capture--recapture models allowing for misidentification
and individual heterogeneity in parameters. Our approach is
computationally more demanding than the closed population
misidentification model proposed by \citet{LukacsBurnham2005},
implemented in Program MARK [\citet{WhiteBurnham1999}], that
allows for individual heterogeneity in detection probability using a
finite mixture distribution. However, \citet{LukacsBurnham2005} do
not properly account for misidentification [\citet
{Yoshizaki2007,LinkEtAl2010,YoshizakiEtAl2011}], and their approach
performs particularly poorly when detection probabilities are too low
($<$0.1) or too high ($>$0.3), as well as when $\alpha<0.95$
[\citet{LukacsBurnham2005}]. The computational cost of our
approach may therefore be worth the additional effort, but similar to
\citet{LinkEtAl2010}, the computational demands of using basis
vectors to propose $\mathbf{x}$ (and allocate $\mathbf{h}$ accordingly) can be
impractical for large $T$. These computational demands can be somewhat
reduced by eliminating basis vectors that will always produce negative
latent history frequencies for a given $\mathbf{f}$, but in the absence of
gains in computing power, more efficient methods for evaluating
equation (\ref{eqLcomplete}) will likely be needed for $T>10$.

Owing to the complexity of the model, we found mixing to be relatively
slow and recommend long runs when implementing our proposed MCMC
algorithm. Other capture--recapture models of somewhat similar
complexity have also exhibited slow mixing that is likely due to
correlated parameters [e.g., \citet
{FienbergEtAl1999,Bonner-Schofield2013,Link2013}] and low movement
rates for the MH sampler [e.g., \citet{LinkEtAl2010}].
Computational efficiency could potentially be improved by accounting
for individual heterogeneity using observed or ``semi-complete'' data
likelihoods in place of complete data likelihoods
[\citet{FienbergEtAl1999,Bonner-Schofield2013}, R.~King, B.~T.~McClintock, D.~Kidney and D.~L.~Borchers, \textit{unpublished
manuscript}]. Capture--recapture
data tend to be somewhat sparse, and in application many of the
possible latent histories could have very low probability. Instead of
drawing basis vectors with equal probability in step~(a) of
the algorithm described in Section~\ref{subsubsecmodel1}, movement
rates of the MH sampler could potentially be improved by drawing $r$
with probabilities proportional to those of the corresponding latent
histories proposed by each basis vector. 

Alternative models have been proposed for handling matching uncertainty
in wildlife populations [e.g., \citet
{WrightEtAl2009,TancrediEtAl2013}] or ``record linkage'' in human
populations [e.g., \citet{Tancredi-Liseo2011}]. These approaches
rely on auxiliary information, such as genotype or family name, to
match recorded histories, but they do not account for individual
heterogeneity in parameters. By integrating a simpler form of record
linkage and individual heterogeneity into a unified missing data
framework, our work constitutes a step toward the ``grand synthesis''
identified by \citet{Fienberg-ManriqueVallier2009} in the context
of multiple recapture estimation, but further development is needed to
integrate auxiliary information into the matching process.

To facilitate Bayesian inference using our approach, we extended
standard probit regression techniques to latent multinomial models
where both the dimension and zeros of the response are unobserved due
to imperfect detection and misidentification. We note that in the
absence of misidentification (i.e., $\alpha=1$), our probit model
provides a convenient Gibbs sampler for Bayesian analysis of
traditional closed population capture--recapture data with heterogeneous
detection probabilities. By avoiding the need to tune proposal
distributions, our probit formulation is potentially a more efficient
alternative to traditional capture--capture models that rely on the
logit link function to account for variability in detection probability
or other parameters. However, we note that the logit link is sometimes
desirable due to its ease of interpretation of the resulting
odds-ratio; recent work by \citet{PolsonEtAl2013} could
potentially be adapted to yield a Gibbs sampler for capture--recapture
models using the logit link.

Capture--recapture models are more robust to individual capture
heterogeneity when absolute abundance is not the focal parameter [e.g.,
\citet{WilliamsEtAl2002}]. In this case, it may be more sensible
to focus on individual heterogeneity in demographic parameters, such as
survival probability [e.g., \citet{Royle2008,GimenezChoqet2010}].
A similarly-structured MCMC algorithm to those described in
Sections~\ref{subsubsecmodel1} and \ref{subsubsecmodel2} can be
employed for other capture--recapture models extended for
misidentification, including open population models, such as the
Cormack--Jolly--Seber (CJS) and more recent multi-state formulations
[e.g., \citet{Pradel2005,MorrisonEtAl2011}]. This is accomplished
by substituting the desired form for the likelihood $[{\mathbf h} \mid\bolds
{\theta},{\bolds\rho}]$ in equation (\ref{eqposterior}) and assigning
corresponding priors for $\bolds{\theta}$ and $\bolds{\rho}$. The proposal
density $ [ \mathbf{h}^* \mid\mathbf{h}, \bolds{\theta}, \bolds{\rho}  ]$
and the set of basis vectors $ \{ \mathbf{v}  \}$ used to update
$\mathbf{h}$ and $\mathbf{x}$, respectively, will depend on the particular
model and the relationship between recorded and latent histories
(formally described by $\mathbf{A}$). 

\citet{BonnerHolmberg2013} and \citet{McClintockEtAl2013a}
recently developed methods for integrated analyses of multiple sources
of capture--recapture data, such as those arising from photo and DNA
records. The methods developed in this paper for incorporating
parameter heterogeneity could be extended to these latent multinomial
models as well. Covariates explaining individual heterogeneity in
parameters [e.g., \citet{KingEtAl2008}] could also be
accommodated. While we have generalized the approach of \citet
{LinkEtAl2010} to a broader suite of misidentification models, we have
maintained several key assumptions that may not be reasonable for many
passive sampling data sets (e.g., those based on visual sightings).
Some challenging (but needed) extensions include the evolving mark
problem examined by \citet{YoshizakiEtAl2009}, allowing for ghost
histories to consist of multiple encounters, and allowing
identification errors to match legitimate individuals.

\section*{Acknowledgments}\label{Acknowledgements}
D. Etter and P. Lukacs for helpful discussions and assistance with the
bear data. K. Scribner and S. Winterstein for pivotal roles in the bear
study. D. Johnson for helpful discussions. The findings and
conclusions in the paper are those of the author(s) and do not
necessarily represent the views of the National Marine Fisheries
Service, NOAA. Any use of trade, product or firm names does not imply
an endorsement by the US Government.



%

\printaddresses
\end{document}